\documentstyle[amssymb,floats,aps,prb,epsf]{revtex}
\epsfclipon 
\begin{document}
\twocolumn[\hsize\textwidth\columnwidth\hsize\csname
@twocolumnfalse\endcsname 

\draft

\title{Anomalous c-axis charge dynamics in copper oxide materials}
\author{Shiping Feng}
\address{CCAST (World Laboratory) P. O. Box 8730, Beijing 100080,
China and \\
$^{*}$Department of Physics, Beijing Normal University, Beijing
100875, China and \\
National Laboratory of Superconductivity, Academia Sinica,
Beijing 100080, China}
\author{Feng Yuan, Weiqiang Yu, and Pengpeng Zhang}
\address{Department of Physics, Beijing Normal University, Beijing
100875, China}

\maketitle

\begin{abstract}
Within the $t$-$J$ model, the c-axis charge dynamics of the copper
oxide materials in the underdoped and optimally doped regimes is
studied by considering the incoherent interlayer hopping. It is
shown that the c-axis charge dynamics is mainly governed by the
scattering from the in-plane fluctuations. In the optimally doped
regime, the c-axis resistivity is a linear in temperatures, and
shows the metallic-like behavior for all temperatures, while the
c-axis resistivity in the underdoped regime is characterized by a
crossover from the high temperature metallic-like to the low
temperature semiconducting-like behavior, which are consistent with
experiments and numerical simulations.
\end{abstract}
]
\narrowtext

The copper oxide materials are among the most complex systems studied
in condensed matter physics. The complications arise mainly from (1)
strong anisotropy in the properties parallel and perpendicular to the
CuO$_{2}$ planes which are the key structural element in the whole
copper oxide superconducting materials, and (2) extreme sensitivity of
the properties to the compositions (stoichiometry) which control the
carrier density in the CuO$_{2}$ plane \cite{n1}. After over ten years
of intense experimental study of the copper oxide materials, a
significant body of reliable and reproducible data has been accumulated
by using many probes, which indicates that the normal-state properties
in the underdoped and optimally doped regimes are quite unusual in many
aspects suggesting the unconventional metallic state realized \cite{n2}.
Among the striking features of the normal-state, the quantity which
most evidently displays the anisotropic property in the copper oxide
materials is the charge dynamics \cite{n3}, which is manifested by the
optical conductivity and resistivity. It has been shown from the
experiments that the in-plane charge dynamics is rather universal within
the whole copper oxide materials \cite{n2,n3}. The in-plane optical
conductivity for the same doping is nearly materials independent both
in the magnitude and energy dependence, and shows the non-Drude
behavior at low energies and anomalous midinfrared band in the
charge-transfer gap, while the in-plane resistivity $\rho_{ab}(T)$
exhibits a linear behavior in the temperature in the optimally doped
regime and a nearly temperature linear dependence with deviations at
low temperatures in the underdoped regime \cite{n2,n3,n4}. By
contrast, the magnitude of the c-axis charge dynamics in the
underdoped and optimally doped regimes is strongly materials
dependent, {\it i.e.}, it is dependent on the species of the building
blocks in between the CuO$_{2}$ planes \cite{n5}. Although the c-axis
charge dynamics is very complicated, some qualitative features seem
to be common, such as (1) for the optimally doped
YBa$_{2}$Cu$_{3}$O$_{6+x}$ and overdoped La$_{2-x}$Sr$_{x}$CuO$_{4}$
systems \cite{n6,n7}, the transferred weight in the c-axis
conductivity decaies as $\rightarrow 1/\omega$ at low energies, which
is in accordance with the metallic-like c-axis resistivity
$\rho_{c}(T)$ for all temperatures, and (2) for the lower doping the
temperature dependent c-axis resistivity $\rho_{c}(T)$ is
characterized by a crossover from the high temperature metallic-like
behavior to the low temperature semiconducting-like behavior
\cite{n5,n6,n7,n8}. The nature of the c-axis charge dynamics in the
copper oxide materials is of great importance, as the
superconducting mechanism is closely associated with the anisotropic
normal-state properties \cite{n9}.

Since the undoped copper oxide materials are antiferromagnetic Mott
insulators, upon hole doping, the antiferromagnetic long-range-order
is rapidly destroyed and the unusual metallic state emerges
\cite{n10}. In this case, many researchers believe that the
essential physics is contained in the doped antiferromagnets
\cite{n11,n12}, which may be effectively described by the $t$-$J$
model acting on the space with no doubly occupied sites
\cite{n11,n12}. On the other hand, there is a lot of evidence from
the experiments and numerical simulations in favour of the $t$-$J$
model as the basic underlying microscopic model \cite{n10,n13}.
Within the two-dimensional (2D) $t$-$J$ model, the in-plane charge
dynamic of the copper oxide materials has been extensively studied
theoretically as well as numerically \cite{n14,n15}. Since the
understanding of the charge dynamics in the copper oxide materials
is not complete without an understanding of the c-axis charge
dynamics, therefore a number of alternative mechanisms for the c-axis
charge dynamics have been proposed \cite{n151,n152}, and the most
reliable results for the c-axis charge dynamics have been obtained by
the numerical simulation \cite{n16}. To shed light on this issue, we,
in this paper, apply the fermion-spin approach \cite{n18,n19} to study
the c-axis charge dynamics based on the $t$-$J$ model by considering
the interlayer coupling. Within each CuO$_{2}$ plane, the essential
physics properties are described by the 2D $t$-$J$ model as,
\begin{eqnarray}
H_{l}&=&-t\sum_{i\hat{\eta}\sigma}C^{\dagger}_{li\sigma}
C_{li+\hat{\eta}\sigma} + h.c. + \mu \sum_{i\sigma}
C^{\dagger}_{li\sigma}C_{li\sigma} \nonumber \\
&+& J\sum_{i\hat{\eta}}{\bf S}_{li}\cdot {\bf S}_{li+\hat{\eta}} ,
\end{eqnarray}
where $\hat{\eta}=\pm a_{0}\hat{x},\pm a_{0}\hat{y}$, $a_{0}$ is the
lattice constant of the square planar lattice, which is set as the
unit hereafter, $i$ refers to planar sites within the l-th CuO$_{2}$
plane, $C^{\dagger}_{li\sigma}$ ($C_{li\sigma}$) are the electron
creation (annihilation) operators, ${\bf S}_{li}=C^{\dagger}_{li}
{\bf \sigma}C_{li}/2$ are the spin operators with ${\bf \sigma}=
(\sigma_{x},\sigma_{y},\sigma_{z})$ as the Pauli matrices, and $\mu$
is the chemical potential. The Hamiltonian (1) is supplemented by
the on-site local constraint, $\sum_{\sigma}C^{\dagger}_{li\sigma}
C_{li\sigma}\leq 1$, {\it i.e.}, there be no doubly occupied sites.
For discussing the c-axis charge dynamics, the hopping between
CuO$_{2}$ planes is considered as \cite{n16}
\begin{eqnarray}
H=-t_{c}\sum_{l\hat{\eta}_{c}i\sigma}C^{\dagger}_{li\sigma}
C_{l+\hat{\eta}_{c}i\sigma} + h.c. +\sum_{l}H_{l},
\end{eqnarray}
where $\hat{\eta}_{c}=\pm c_{0}\hat{z}$, $c_{0}$ is the interlayer
distance, and has been determined from the experiments \cite{n17} as
$c_{0} > 2a_{0}$. In the underdoped and optimally doped regimes, the
experimental results \cite{n5,n6,n7,n8} show that the ratio
$R=\rho_{c}(T)/\rho_{ab}(T)$ ranges from $R\sim 100$ to $R >10^{5}$,
this large magnitude of the resistivity anisotropy reflect that the
c-axis mean free path is shorter than the interlayer distance, and
the carriers are tightly confined to the CuO$_{2}$ planes, and also
is the evidence of the incoherent charge dynamics in the c-axis
direction. Therefore the c-axis momentum can not be defined
\cite{n21}. Moreover, the absence of the coherent c-axis charge
dynamics is a consequence of the weak interlayer hopping matrix
element $t_{c}$, but also of a strong intralayer scattering,
{\it i.e.}, $t_{c}\ll t$, and therefore the common CuO$_{2}$ planes
in the copper oxide materials clearly dominate the most
normal-state properties. In this case, the most relevant for the
study of the c-axis charge dynamics is the results on the in-plane
conductivity $\sigma_{ab}(\omega)$ and related single-particle
spectral function $A(k,\omega)$.

Based on the 2D $t$-$J$ model, the self-consistent mean-field theory
in the underdoped and optimally doped regimes has been developed
\cite{n19} within the fermion-spin approach \cite{n18}, which has
been applied to study the photoemission, electron dispersion and
electron density of states in the copper oxide materials, and the
results are qualitative consistent with the experiments and
numerical simulations. Moreover, the in-plane charge dynamics in
the copper oxide materials has been discussed \cite{n14} by
considering the fluctuations around this mean-field solution, and
the results exhibits a behavior similar to that seen in the
experiments \cite{n4} and numerical simulations \cite{n15}. In the
fermion-spin theory \cite{n18,n19}, the constrained electron
operators in the $t$-$J$ model is decomposed as,
\begin{eqnarray}
C_{li\uparrow}=h^{\dagger}_{li}S^{-}_{li}, ~~~~~~ C_{li\downarrow}
=h^{\dagger}_{li}S^{+}_{li},
\end{eqnarray}
with the spinless fermion operator $h_{i}$ keeps track of the
charge (holon), while the pseudospin operator $S_{i}$ keeps track
of the spin (spinon), then it naturally incorporates the physics of
the charge-spin separation. The main advantage of this approach is
that the electron on-site local constraint can be treated exactly
in analytical calculations. In this case, the low-energy behavior
of the $t$-$J$ model (2) in the fermion-spin representation can be
rewritten as \cite{n14},
\begin{mathletters}
\begin{eqnarray}
H &=& t_{c}\sum_{l\hat{\eta}_{c}i}h^{\dagger}_{l+\hat{\eta}_{c}i}
h_{li}(S^{+}_{li}S^{-}_{l+\hat{\eta}_{c}i}+S^{-}_{li}
S^{+}_{l+\hat{\eta}_{c}i})+ \sum_{l}H_{l}, \\
H_{l} &=& t\sum_{i\hat{\eta}}h^{\dagger}_{li+\hat{\eta}}h_{li}
(S^{+}_{li}S^{-}_{li+\hat{\eta}}+S^{-}_{li}S^{+}_{li+\hat{\eta}})
- \mu \sum_{i}h^{\dagger}_{li}h_{li} \nonumber \\
&+& J_{eff}\sum_{i\hat{\eta}}({\bf S}_{li}\cdot
{\bf S}_{li+\hat{\eta}}),
\end{eqnarray}
\end{mathletters}
where $J_{eff}=J[(1-\delta)^{2}-\phi ^{2}]$, the holon particle-hole
parameter $\phi=\langle h^{\dagger}_{li}h_{li+\hat{\eta}}\rangle$,
and $S^{+}_{li}$ and $S^{-}_{li}$ are the pseudospin raising and
lowering operators, respectively. These pseudospin operators obey
the Pauli algebra, {\it i.e.}, they behave as fermions on the same
site, and as bosons on different sites. It is shown \cite{n18} that
the constrained electron operator in the $t$-$J$ model can be mapped
exactly onto the fermion-spin transformation defined with an
additional projection operator. However, this projection operator is
cumbersome to handle for the actual calculation possible in 2D, we
have dropped it in Eq. (4). It has been shown in Ref. \cite{n18}
that such treatment leads to the errors of the order $\delta$ in
counting the number of spin states, which is negligible for small
doping $\delta$.

In the framework of the charge-spin separation, an electron is
represented by the product of a holon and a spinon, then the
external field can only be coupled to one of them. Ioffe and
Larkin \cite{n20} and Li {\it et al}. \cite{n22} have shown that
the physical conductivity $\sigma(\omega)$ is given by,
\begin{eqnarray}
\sigma^{-1}(\omega)=\sigma^{(h)-1}(\omega)+\sigma^{(s)-1}(\omega),
\end{eqnarray}
where $\sigma^{(h)}(\omega)$ and $\sigma^{(s)}(\omega)$ are the
contributions to the conductivity from holons and spinons,
respectively. Within the Hamiltonian (4), the c-axis current
densities of holons and spinons are given by the time derivative
of the polarization operator using Heisenberg's equation of motion
as, $j^{(h)}_{c}=2\tilde{t}_{c}e\chi\sum_{l\hat{\eta}_{c}i}
\hat{\eta}_{c}h_{l+\hat{\eta}_{c}i}^{\dagger} h_{li}$ and
$j^{(s)}_{c}=t_{c}e\phi_{c}\sum_{l\hat{\eta}_{c}i}\hat{\eta}_{c}
(S^{+}_{li}S^{-}_{l+\hat{\eta}_{c}i}+S^{-}_{li}
S^{+}_{l+\hat{\eta}_{c}i})$, respectively, where $\tilde{t}_{c}=
t_{c}\chi_{c}/\chi$ is the effective interlayer holon hopping
matrix element, and the mean-field spinon and holon order
parameters are defined \cite{n19} as $\chi_{c}=\langle S_{li}^{+}
S_{l+\hat{\eta}_{c}i}^{-}\rangle$, $\chi=\langle S_{li}^{+}
S_{li+\hat{\eta}}^{-}\rangle$, and $\phi_{c}=\langle
h^{\dagger}_{li}h_{l+\hat{\eta}_{c}i}\rangle$. As in the previous
discussions \cite{n14}, a formal calculation for the spinon part
shows that there is no the direct contribution to the
current-current correlation from spinons, but the strongly
correlation between holons and spinons is considered through the
spinon's order parameters entering in the holon part of the
contribution to the current-current correlation, therefore the
charge dynamics in the copper oxide materials is mainly caused
by the charged holons within the CuO$_{2}$ planes, which are
strongly renormalized because of the strong interactions with
fluctuations of the surrounding spinon excitations. In this case,
the c-axis optical conductivity is expressed \cite{n23} as
$\sigma_{c}(\omega)=-{\rm Im}\Pi^{(h)}_{c}(\omega)/\omega$ with
the c-axis holon current-current correlation function
$\Pi^{(h)}_{c}(t-t')=\langle\langle j^{(h)}_{c}(t)j^{(h)}_{c}
(t')\rangle\rangle$. In the case of the incoherent charge
dynamics in the c-axis direction, {\it i.e.}, the weak interlayer
hopping $t_{c}\ll t$, this c-axis holon current-current
correlation function $\Pi^{(h)}_{c}(\omega)$ can be evaluated in
terms of the in-plane holon Green's function $g(k,\omega)$
\cite{n14,n16}, then we obtain the c-axis optical conductivity as
\cite{n14,n16,n24},
\begin{eqnarray}
\sigma_{c}(\omega)&=&{1\over 2}(4\tilde{t}_{c}e\chi c_{0})^2
{1\over N}\sum_k\int^{\infty}_{-\infty}{d\omega'\over 2\pi}
A_{h}(k,\omega'+\omega) \nonumber \\
&\times& A_{h}(k,\omega'){n_{F}(\omega'+\omega)-n_{F}(\omega')
\over \omega},
\end{eqnarray}
where $n_{F}(\omega)$ is the Fermi distribution functions, the
in-plane holon spectral function $A_{h}(k,\omega)=-2{\rm Im}g(k,
\omega)$, while the in-plane holon Green's function $g(k,\omega)$
has been obtained by considering the second-order correction for
holons due to the antiferromagnetic fluctuations, and given in Ref.
\cite{n14}. As pointed in Ref. \cite{n16}, the approximation
assumption of the independent electron propagation in each layer
has been used in the above discussions, and is justified for
$t_{c}\ll t$, therefore the c-axis conductivity is essentially
determined by the properties of the in-plane spectral function.
We have performed a numerical calculation for the c-axis optical
conductivity $\sigma_{c}(\omega)$, and the results at the doping
$\delta=0.12$ (solid line), $\delta=0.09$ (dashed line), and
$\delta=0.06$ (dot-dashed line) for the parameters $t/J=2.5$,
$\tilde{t}_{c}/t=0.04$, and $c_{0}/a_{0}=2.5$  at the temperature
$T$=0 are shown in Fig. 1, where the charge $e$ has been set as
the unit. From Fig. 1, it is found that $\sigma_{c}(\omega)$ is
composed of two bands separated at $\omega \sim 0.4t$, the
higher-energy band, corresponding to the "midinfrared band" in
the in-plane optical conductivity $\sigma_{ab}(\omega)$
\cite{n4,n14}, shows a broad peak at $\omega\sim 0.7t$, moreover,
the weight of this band is strongly doping dependent, and
decreasing rapidly with dopings, but the peak position does not
appreciably shift to higher energies, which is consistent with the
experimental results \cite{n5,n6}. On the other hand, the
transferred weight of the lower-energy band forms a sharp peak at
$\omega<0.4t$, which can be described formally by the non-Drude
formula, and our analysis indicates that this peak decay is
$\rightarrow 1/\omega$ at low energies as in the case of
$\sigma_{ab}(\omega)$ \cite{n4,n14}. In comparison with
$\sigma_{ab}(\omega)$ \cite{n14}, the present results also show
that the values of $\sigma_{c}(\omega)$ are by $2\sim 3$ orders of
magnitude smaller than those of $\sigma_{ab}(\omega)$ in the
corresponding energy range. For further understanding the property
of $\sigma_{c}(\omega)$, we have also discussed the finite
temperature behavior of $\sigma_{c}(\omega)$, and the numerical
results at the doping $\delta=0.12$ for the parameters $t/J=2.5$,
$\tilde{t}_{c}/t=0.04$, and $c_{0}/a_{0}=2.5$ with $T=0.2J$ (solid
line) and $T=0.5J$ (dashed line) are plotted in Fig. 2, which show
that $\sigma_{c}(\omega)$ is temperature dependent for
$\omega<1.2t$ and almost temperature independent for $\omega>1.2t$,
while the higher-energy band is severely suppressed with increasing
temperatures, and vanishes at higher temperature ($T>0.4J$). These
results are also qualitative consistent with the experimental
results \cite{n5,n6} and numerical simulations \cite{n16}.

\begin{figure}[prb]
\epsfxsize=3.0in\centerline{\epsffile{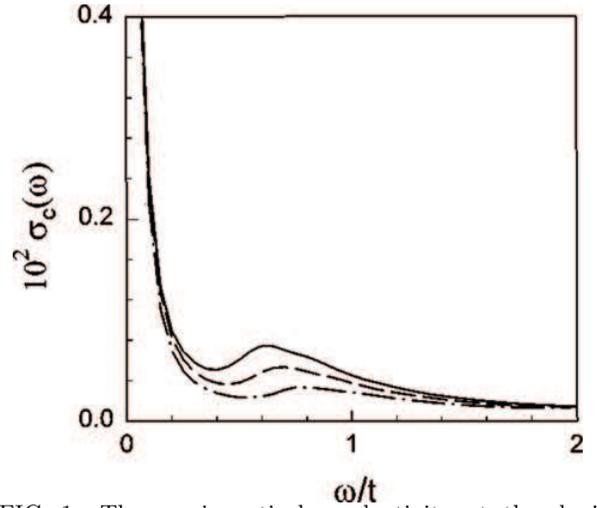}}
\caption{The c-axis optical conductivity at the doping
$\delta=0.12$ (solid line),  $\delta=0.09$ (dashed line), and
$\delta=0.06$ (dot-dashed line) for $t/J=2.5$,
$\tilde{t}_{c}/t=0.04$, and $c_{0}/a_{0}=2.5$ with the
temperature $T=0$.}
\end{figure}

\begin{figure}[prb]
\epsfxsize=3.0in\centerline{\epsffile{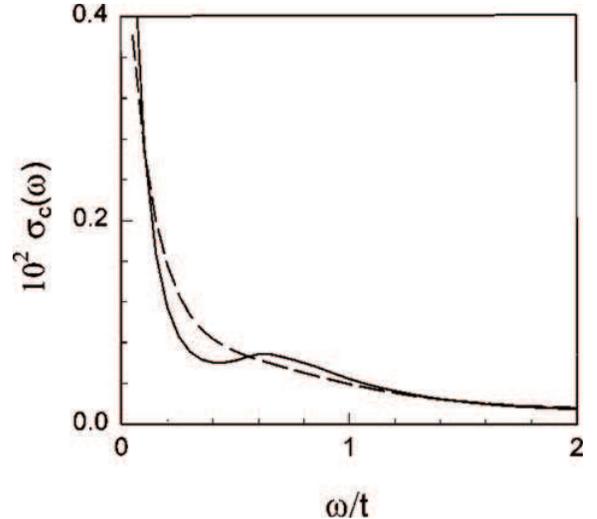}}
\caption{The c-axis optical conductivity at the doping $\delta=0.12$
for $t/J=2.5$, $\tilde{t}_{c}/t=0.04$, and $c_{0}/a_{0}=2.5$ with
the temperature $T=0.2J$ (solid line) and $T=0.5J$ (dashed line).}
\end{figure}

\begin{figure}[prb]
\epsfxsize=3.0in\centerline{\epsffile{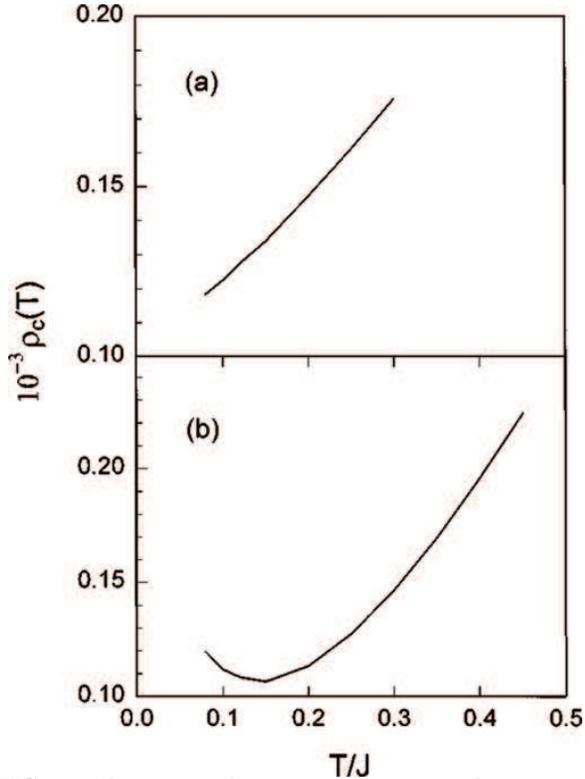}}
\caption{The c-axis electron resistivity at the parameter $t/J=2.5$,
$\tilde{t}_{c}/t=0.04$, and $c_{0}/a_{0}=2.5$ for (a) the doping
$\delta=0.12$ and (b) $\delta=0.06$.}
\end{figure}

The quantity which is closely related with the c-axis
conductivity is the c-axis resistivity $\rho_{c}(T)$, and can
be expressed as,
\begin{eqnarray}
\rho_{c}={1\over \lim_{\omega\rightarrow 0} \sigma_{c}(\omega)}.
\end{eqnarray}
This c-axis resistivity has been evaluated numerically and the
results at the doping $\delta=0.12$ and $\delta=0.06$ for the
parameters $t/J=2.5$, $\tilde{t}_{c}/t=0.04$, and $c_{0}/a_{0}=2.5$
are shown in Fig. 3(a) and Fig. 3(b), respectively. In the
underdoped regime, the behavior of the temperature dependence of
$\rho_{c}(T)$ shows a crossover from the high temperature
metallic-like ($d\rho_{c}(T)/dT>0$) to the low temperature
semiconducting-like ($d\rho_{c}(T)/dT<0$), but the metallic-like
temperature dependence dominates over a wide temperature range. In
comparison with the in-plane resistivity $\rho_{ab}(T)$ \cite{n14},
it is shown that the crossover to the semiconducting-like range in
$\rho_{c}(T)$ is obviously linked with the crossover from the
temperature linear to the nonlinear range in $\rho_{ab}(T)$, and
are caused by the pseudogap observed in the normal-state, but
$\rho_{ab}(T)$ is only slightly affected by the pseudogap \cite{n14},
while $\rho_{c}(T)$ is more sensitive to the underlying mechanism.
Our results also show that there is the common origin for these
crossovers. Therefore in this case, there is a general trend that
the copper oxide materials show nonmetallic $\rho_{c}(T)$ in the
underdoped regime at low temperatures. While in the optimally doped
regime, $\rho_{c}(T)$ is a linear in temperatures, and shows the
metallic-like behavior for all temperatures. These results are
qualitative consistent with the experimental results
\cite{n5,n6,n7,n8} and numerical simulation \cite{n16}. It has been
shown from the experiments \cite{n241} that the charge dynamics in
some strongly correlated ladder materials shows the similar
behaviors.

In the above discussions, the central concerns of the c-axis charge
dynamics in the copper oxide materials are the two dimensionality
of the electron state and incoherent hopping between the CuO$_{2}$
planes, and therefore the c-axis charge dynamics in the present
fermion-spin picture is determined by the in-plane charged holon
fluctuations. In this case, the c-axis scattering rate is associated
with the in-plane scattering rate, and can be roughly described by
the imaginary part of the self-energy of the charged holons within
the CuO$_{2}$ planes, which is consistent with the "dynamical
dephasing" theory proposed by Leggett \cite{n151}.

In the fermion-spin theory \cite{n18}, the charge and spin degrees
of freedom of the physical electron are separated as the holon and
spinon, respectively. Although both holons and spinons contributed
to the charge and spin dynamics, but it has been shown that the
scattering of spinons dominates the spin dynamics \cite{n25}, while
the results of the in-plane charge dynamics \cite{n14} and present
c-axis charge dynamics shows that scattering of holons dominates
the charge dynamics, therefore the notion of the charge-spin
separation naturally accounts for all the qualitative features of
the normal-state properties of the copper oxide materials. To our
present understanding, the main reasons why the fermion-spin theory
based on the charge-spin separation is successful in studying the
normal-state property of the strongly correlated copper oxide
materials are that (1) the electron single occupancy on-site local
constraint is exactly satisfied in the analytic calculation. Since
the anomalous normal-state property of the copper oxide materials
are caused by the strong electron correlation in these systems
\cite{n10,n13}, and can be effectively described by the $t$-$J$
model \cite{n10,n11,n12,n13}, but the strong electron correlation
in the $t$-$J$ model manifests itself by the electron single
occupancy on-site local constraint, then the satisfaction of this
local constraint is equivalent to that the strong electron-electron
interaction has been properly treated. This is why the crucial
requirement is to treat this constraint exactly in the $t$-$J$ model
in the analytic discussions. (2) Since the local constraint is
satisfied even in the mean-field approximation within the
fermion-spin theory \cite{n18}, the extra gauge degree of freedom
related to the common "flux" phase problem occurring in the
slave-particle approach \cite{n26} does not appear here, which is
confirmed by our previous discussions within the mean-field theory
\cite{n19}, where the photoemission, electron dispersion and
electron density of states in the copper oxide materials have been
studied, and the results are qualitative similar
to that seen in the experiments and numerical simulations. (3) As
mentioned above, the dropping the projection operator in Eq. (4)
will only lead to errors of the order $\delta$ in counting the
number of spin states within the common decoupling approximation
\cite{n26}. This because that the constrained electron operators
$C_{i\sigma}$ in the $t$-$J$ model can be also mapped onto the
slave-fermion formulism \cite{n26} as $C_{i\sigma}=h^{\dagger}_{i}
a_{i\sigma}$ with the local constraint $h^{\dagger}_{i}h_{i}+
\sum_{\sigma}a^{\dagger}_{i\sigma}a_{i\sigma}=1$. We can solve
this constraint by rewritting the boson operators $a_{i\sigma}$
in terms of the CP$^{1}$ boson operators $b_{i\sigma}$ as
$a_{i\sigma}=(1-h^{\dagger}_{i}h_{i})^{1/2}b_{i\sigma}\approx
(1-h^{\dagger}_{i}h_{i}/2)b_{i\sigma}$ supplemented by the local
constraint $\sum_{\sigma}b^{\dagger}_{i\sigma}b_{i\sigma}=1$.
Since the CP$^{1}$ boson operators $b_{i\uparrow}$ and
$b_{i\downarrow}$ with the local constraint can be identified with
the pseudospin lowering and raising operators in the fermion-spin
approach \cite{n18}, respectively, then the spinon propagator in
the restricted Hilbert space can be written as $D_{R}(i-j,t-t')=
\langle\langle[1-h^{\dagger}_{i}(t)h_{i}(t)/2];[1-h^{\dagger}_{i}
(t')h_{i}(t')/2]\rangle\rangle D(i-j,t-t')\approx [1-\delta-
O(\delta^{2})]D(i-j,t-t')$, where $D(i-j,t-t')$ is the spinon
propagator within the fermion-spin approach. In this case, the
extra spin degrees of freedom in the fermion-spin theory only lead
to the errors of the order $\delta$ in calculating the spinon
propagator within the common decoupling approximation \cite{n26},
which is negligible for small doping $\delta$. This is why the
theoretical results of the spin dynamics within the fermion-spin
approach \cite{n25} are qualitative consistent with the experiments
and numerical simulations.

In summary, we have studied the c-axis charge dynamics of the
copper oxide materials within the $t$-$J$ model by considering the
incoherent interlayer hopping. Our results show that the c-axis
charge dynamics is mainly governed by the scattering from the
in-plane charged holon fluctuations. The c-axis optical
conductivity and resistivity have been discussed, and the results
are qualitative consistent with the experiments and numerical
simulations.

Finally we also note that since the structure of the building
blocks in between the CuO$_{2}$ planes for the chain copper
oxide materials is different from these for the nochain copper
oxide materials, some subtle differences between the chain and
nochain copper oxide materials for the c-axis charge dynamics
have been found from the experiments \cite{n5,n6}. It has been
suggested \cite{n8} that for the nochain copper oxide materials
the doped holes may introduce the disorder in between the
CuO$_{2}$ planes, contrary to the case of the chain copper oxide
materials, where the increasing doping reduces the disorder in
between the CuO$_{2}$ planes. It is possible that the disorder
introduced by the doped holes residing between layers in the
nochain copper oxide materials in the underdoped regime may modify
the interlayer hopping elements, which leads to the subtle
differences between the chain and nochain copper oxide materials
for the c-axis charge dynamics. These and other related issues
are under investigation now.

\acknowledgments
The authors would like to thank Professor Ru-Shan Han and Professor
H. Q. Lin for helpful discussions. This work was supported by the
National Natural Science Foundation under Grant No. 19774014 and
the State Education Department of China through the Foundation of
Doctoral Training. The partial support from the Earmarked Grant for
Research from the Research Grants Council of Hong Kong, China are
also acknowledged.

\end{document}